\documentclass[proof]{WileyASNA-v1}

\articletype{Article Type}%

\begin{document}

\title{Possible optical counterparts of ULXs in NGC 1672}

\author[1,3]{S. Allak*}

\author[2,3]{A. Akyuz}

\author[4]{E. Sonbas}

\author[5]{K. S. Dhuga}

\authormark{S. Allak \textsc{et al}}

\address[1]{\orgdiv{Department of Physics, University of Çanakkale Onsekiz Mart, 17100, Çanakkale, Türkiye}}

\address[2]{\orgdiv{Department of Physics, University of Çukurova, 01330, Adana, Türkiye}}

\address[3]{\orgdiv{Space Science and Solar Energy Research and Application Center (UZAYMER), University of Çukurova, 01330, Adana, Türkiye}}
\address[4]{\orgdiv{Adiyaman University, Department of Physics, 02040 Adiyaman, Türkiye}}

\address[5]{\orgdiv{Department of Physics, The George Washington University, Washington, DC 20052, USA}}

\corres{* \email{0417allaksinan@gmail.com}}

\abstract{In this study, we use archival data from {\it HST}, {\it Chandra}, {\it XMM-Newton}, and {\it Swift-XRT}, to probe the nature of 9 (X1-X9) candidate ULXs in NGC 1672.
 Our study focuses on using the precise source positions obtained via improved astrometry based on {\it Chandra} and {\it HST} observations to search for and identify optical counterparts for these ULXs.
 Unique optical counterparts are identified for X2 and X6; two potential counterparts were determined for X1, X5 and X7 within the respective error radii while no optical counterparts were found for the remaining four sources.
Based on spectral energy distributions (SEDs), X-ray and optical temporal analyses, some evidences about the nature of X1 and X2 were obtained.}


\maketitle

\section{Introduction}

Ultraluminous X-ray sources (ULXs) are non-nuclear point-like sources in a number of external galaxies. Their X-ray luminosities are above a threshold luminosity of L$_{X}$ > 10$^{39}$ erg s$^{-1}$, exceeding the Eddington limit for a typical 10 M$_{\odot}$ stellar-remnant black hole (see review by \citealp{2017ARA&A..55..303K, 2021AstBu..76....6F}).

An early model, though now seemingly less likely, poses the existence of BHs in the intermediate mass range of M $\sim$ 10$^{2}$- 10$^{5}$ M$_{\odot}$, accreting at sub-Eddington rates \citep{1999ApJ...519...89C,2004ApJ...614L.117M}. Based on recently studies, tend to lean toward stellar-mass compact objects with a possible
combination of effects such as geometric beaming, and/or accretion at super-Eddington limits \citep{2002ApJ...568L..97B,2007Ap&SS.311..203R, 2007MNRAS.377.1187P,2009MNRAS.393L..41K,2018ApJ...857L...3W}. Indeed, the recent detection of pulsations in a handful of ULX sources strongly argues in favor of at least a fraction of these sources hosting neutron stars (NSs) \citep{2014Natur.514..202B,2016ApJ...831L..14F,2017Sci...355..817I,2017MNRAS.466L..48I,2018MNRAS.476L..45C,2019MNRAS.488L..35S,2020ApJ...895...60R}. Likewise, the presence of a cyclotron resonance scattering feature in the X-ray spectrum of ULX-8 in M51, discovered by \cite{2018NatAs...2..312B}, provides strong evidence for the nature of the compact object i.e., a NS, and a measure of the associated magnetic field.

In the present work, we focus on NGC 1672, a distance of 16.3 Mpc \citep{2000AJ....119..612D}. This galaxy has been studied by \cite{2011ApJ...734...33J}, who identified 9 ULX candidates. Our primary goal is to search for and identify potential optical counterparts for these ULXs. In addition, a secondary goal is to use the available {\it Chandra}, {\it XMM-Newton} and {\it Swift} archival data, to study the X-ray spectral and temporal variations of the candidate ULXs.

Optical studies provide valuable information regarding the nature of the donor star, disk geometry, and can place constraints on the mass of the accretor. Technically, the optical emission observed in ULX binaries can be due to the accretion disk, the donor star, and/or some combination of both. Many recent studies \citep{2007MNRAS.376.1407C,2008MNRAS.386..543P,2011ApJ...737...81T,2012ApJ...745..123G,2012MNRAS.420.3599S,2014MNRAS.444.2415S,2018MNRAS.480.4918A,2019ApJ...884L...3Y}, focusing on optical variability, multi-band colors, and SED modeling, strongly suggest that the optical emission is likely contaminated or even dominated by reprocessed radiation from an irradiated accretion disk.

Identification of point-like (potential) counterparts of ULXs with blue colors are indicative of early-type, OB stars. \citep{2010MNRAS.403L..69P,2012ApJ...758...28J,2018ApJ...854..176V}. As a rare example, \cite{2014Natur.514..198M}
reported that photospheric absorption lines have been detected from the donor star in the blue part of the spectrum of P13 in NGC 7793. On the other hand, nearby ULX counterparts in the near-infrared might be red supergiants \citep{2016MNRAS.459..771H,2020MNRAS.497..917L}.

\section{Data Reduction and Methodology} \label{sec:2}
\subsection{Optical} 

NGC 1672 was observed by {\it HST} in 2005 and 2019 using the ACS/WFC (Advanced Camera for Surveys/Wide Field Channel) and WFC3/UVIS (The Wide Field Camera 3), respectively.

\begin{figure}
\begin{center}
\includegraphics[width=\columnwidth]{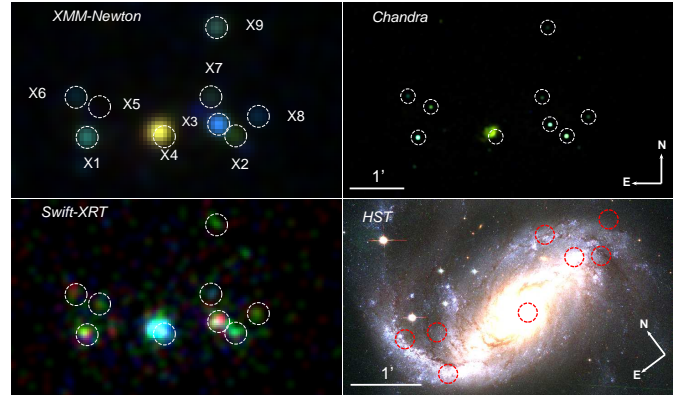}
\caption{Three-color X-ray (upper left {\it XMM-Newton}, upper right {\it Chandra} and lower left {\it Swift-XRT}) and {\it HST} (lower right) images of the NGC 1672 galaxy. For X-ray images, red, green and blue (RGB) represent 0.3–1 keV, 1–2.5 keV and 2.5–8 keV emissions, respectively and images smoothed with a 5 arcsec Gaussian (X4 is distinct from the nucleus). The ULX candidates are indicated with dashed white circles. Red circles represent the positions of 8 ULXs on the {\it HST} RGB image (R:F814W, G:F550M and B:F435W). All X-ray images are the same scale.}
\label{F:1}
\end{center}
\end{figure}

We note at the outset that our numbering scheme for the sources differs from that used by \cite{2011ApJ...734...33J}; we label the candidate ULXs according to their increasing {\it Chandra} counts i.e., X1 (\#25), X2 (\#4), X3 (\#5), X4 (\#13), X5 (\#24), X6 (\#28), X7 (\#7), X8 (\#1) and X9 (\#6). The numbers in parentheses correspond to the source numbers in \cite{2011ApJ...734...33J}. Figure \ref{F:1} shows the location of the 9 ULX candidates on three-color images.

{\it Chandra} and {\it HST} images were used to obtain astrometric precision by following the method we used in our previous studies \cite{2022MNRAS.510.4355A} (and references therein). We chose deep {\it Chandra} ACIS-S (ObsID 5932) and {\it HST} ACS (ObsID j6n202010) images for astrometric corrections of the sources. As a result, we found the positions of the optical counterparts of ULXs on the {\it HST} image within an error radius of 0.21 arcsec at 95\% confidence level.

\begin{figure}
\begin{center}
\includegraphics[width=\columnwidth]{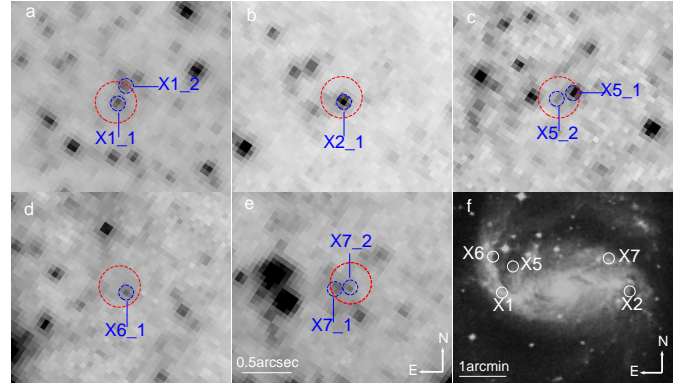}
\caption{The identified optical counterparts of the five ULXs are shown on the F814W images. Red dashed circles indicate the astrometric error radius of the optical counterparts. Blue dashed circles represent the center coordinates of each optical counterpart. North is up and east is left in all panels. Panels (a-e) are the same scale. In panel f shows positions of ULXs which have optical counterparts on the DSS image.}
\label{F:can}
\end{center}
\end{figure}

We found unique optical counterparts for X2 ({\it X2$\_1$}) and X6 ({\it X6$\_1$}). Two potential counterparts each were found for X1 ({\it {\it X1$\_1$}} and {\it X1$\_2$}), X5 ({\it X5$\_1$} and {\it X5$\_2$} ) and X7 ({\it X7$\_1$} and {\it X7$\_2$}) respectively (Figure \ref{F:can}). However, we could not identify any optical counterpart(s) for X3, X4 and X8 within their respective error radii: X9, unfortunately, was not observed by {\it HST}.

Point-Spread Function (PSF) photometry was performed with {\scshape dolphot} v2.0 \citep{2000PASP..112.1383D} to determine the magnitudes of optical sources using the {\it HST} data. Obtained magnitude and flux values were corrected with Galactic extinction A$_{V}$ = 0.065 mag.\citep{2011ApJ...737..103S}. We calculated the absolute magnitudes, M$_{V}$, for both UVIS/F555W (V) and WFC/F550M ($\simeq$V) with adopted distance of 16.3 Mpc. These values are given in Table \ref{T:tab6}. In addition, the calculated color indices to determine the spectral type of optical counterparts of ULXs are also given in Table \ref{T:tab6}.

\begin{table*}
\centering
\begin{minipage}[b]{0.9\linewidth}
\caption{Properties of optical counterparts}
\begin{tabular}{ccccccccccccc}
\hline
Optical counterparts & M$_{V_{ACS}}$ & M$_{V_{UVIS}}$ & (B-V)$_{0}$ & (V-I)$_{0}$ & ST & log(F$_{X}$/F$_{V}$) & $\alpha_{ox}$ \\
& (1) & (2) & (3) & (4) & (5) & (6) & (7) \\
\hline
{\it X1$\_1$} & -7.13 $\pm$ 0.05 & -6.90 $\pm$ 0.4 & -0.03 & -0.080 & B5-A0 & 51 $\pm$ 6 & -0.69$\pm$0.01 \\
{\it X2$\_1$} & -7.45 $\pm$ 0.04 & -7.48 $\pm$ 0.3 & 0.03 & 0.054 & B9-A3 & 21 $\pm$ 3 & -0.66$\pm$0.01 \\
{\it {\it X5$\_1$}} & -6.21 $\pm$ 0.23 & -5.10 $\pm$ 0.35 & -0.08 & 0.11 & B6-A1 & 13 $\pm$ 2 & -0.71$\pm$0.08 \\
{\it X5$\_2$} & ... & -6.11 $\pm$ 0.16 & -0.47 & ... & ... & ... & -0.55$\pm$0.03 \\
{\it X7$\_1$} & -6.16 $\pm$ 0.12 & -5.34 $\pm$ 0.11 & -0.15 & -0.19 & B2-B6 & 11 $\pm$ 1 & -0.83$\pm$0.02 \\
{\it X7$\_2$} & -5.22 $\pm$ 0.85 & ... & ... & ... & ... & 31 $\pm$ 4 & ... \\
\hline
\end{tabular}
\\ Notes: Absolute magnitudes were obtained (1) from ACS/WFC (2) from ACS/UVIS data with adopted distance 16.3 Mpc \citep{2000AJ....119..612D}. (3) and (4) Color values obtained from F435W-F550M (B-V) and F550M-F814W (V-I) filters, respectively. (5) Spectral types {\it estimated from the color values}. (6) The F$_{X}$/F$_{V}$ ratios. (7) X-ray-UV correlation cast in terms of the so-called optical spectral index ($\alpha_{ox}$). \\
\label{T:tab6}
\end{minipage}
\end{table*}

\begin{figure}
\begin{center}
\includegraphics[width=\columnwidth]{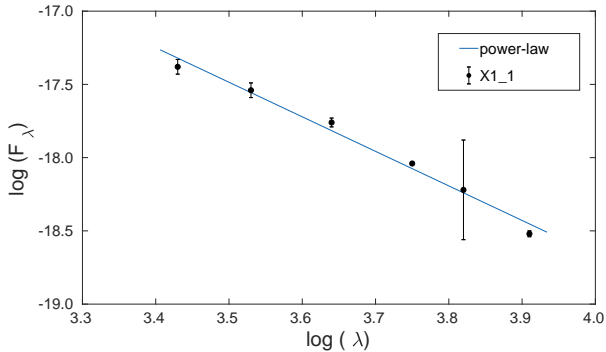}
\caption{The reddening corrected SED of {\it X1$\_1$}. The {\it power-law} model is shown by blue solid line. All data are shown with dark circles and their
respective errors with bars. The {\it power-law} model has $\alpha$ $=$ -2.13 $\pm$ 0.16. The units of y and x axes are erg s$^{-1}$ cm$^{-2}$ \AA$^{-1}$ and \AA, respectively.}
\label{F:SEDsX1}
\end{center}
\end{figure}

\begin{figure}
\begin{center}
\includegraphics[width=\columnwidth]{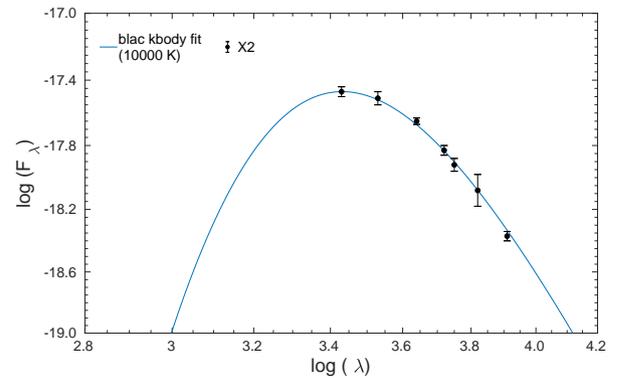}
\caption{The reddening corrected SED of {\it X2$\_1$}. The {\it blackbody} model is shown by blue solid line.}
\label{F:SEDsX2}
\end{center}
\end{figure}

We attempt to constrain the nature of companions by fitting Spectral energy distribution (SEDs) with a {\it blackbody} or a {\it power-law} (F $\propto$ $\lambda^{\alpha}$) spectrum. The SED of {\it X1$\_1$} is adequately fitted by a {\it power-law} spectrum with $\alpha$ $=$ -2.13 $\pm$ 0.16. The {\it X2$\_1$} SED is well fitted by a {\it blackbody} spectrum with a temperature of 10$^{4}$ $\pm$ 113 Kelvin (K) at 95\% confidence level. This temperature corresponds to a late-type B supergiant donor \citep{1981Ap&SS..80..353S}. The reddening corrected SEDs of {\it X1$\_1$} and {\it X2$\_1$} are shown in Figure \ref{F:SEDsX1} and \ref{F:SEDsX2}, respectively. No acceptable model-fits were obtained for the remaining optical counterparts.

In order to investigate the relation of the optical counterparts with their environments and estimate their ages, the color-magnitude diagrams (CMDs) were obtained assuming the optical emission of the donor stars dominate. CMDs were obtained for only X1 and X7 have an association with nearby star groups or clusters.

The ratios of X-ray to optical fluxes (F$_{X}$/F$_{V}$) for the optical counterparts can be used to distinguish between ULXs and AGNs \citep{2020MNRAS.499.2138A} (and references therein). According to a study by \cite{2010MNRAS.401.2531A}, this ratio is defined in the range of 0.1-10 for AGNs (Table \ref{T:tab6}). Moreover, the distinct role played by the accretion disk in the emission process is well known for AGNs \citep{2010A&A...512A..34L, 2013A&A...550A..71V} and is well encapsulated in a X-ray-UV correlation cast in terms of the so-called optical spectral index $\alpha_{ox}$, as defined by \cite{1979ApJ...234L...9T}. For comparison, we extracted $\alpha_{ox}$ for the candidate ULXs in NGC 1672. These $\alpha_{ox}$ values are listed in Table \ref{T:tab6}.

\subsection{X-ray}

NGC 1672 was observed by {\it XMM-Newton} and {\it Chandra} in 2004 and 2006, respectively. In addition, it was observed 17 times by {\it Swift-XRT} between 2006 and 2020.

The {\it Chandra} ACIS (Advanced CCD Imaging Spectrometer) data were analyzed by using {\scshape ciao}\footnote{https://cxc.cfa.harvard.edu/ciao/} v4.12 software with its calibration package {\scshape caldb}\footnote{https://cxc.cfa.harvard.edu/caldb/} v4.9. The {\it XMM-Newton} data reductions were carried out using the {\scshape sas} (Science Analysis Software version 18.0). The {\it Swift-XRT} data sets were processed with HEASoft v6.29\footnote{https://heasarc.gsfc.nasa.gov/docs/software/heasoft/}, the tool {\it xrtpipeline} and calibration files CALDB v4.9. Data used were taken in Photon Counting (PC) mode. We combined the data taken in the same year (e.g. six datasets in 2012) for spectral fitting due to the short exposure time and low data statistics.

\cite{2011ApJ...734...33J} fitted the X-ray spectra with a {\it power-law} model and obtained reasonable fits. In order to explore whether these fits could be improved, the spectra for the ULX candidates which have sufficient statistics were grouped at least 15 counts per energy bin. We then fit the grouped spectra with the following models {\it power-law}, the absorbed multi-color disk blackbody ({\it diskbb}), {\it diskpbb} and {\it cutoffpl}, respectively.

\begin{figure}
\begin{center}
\includegraphics[width=\columnwidth]{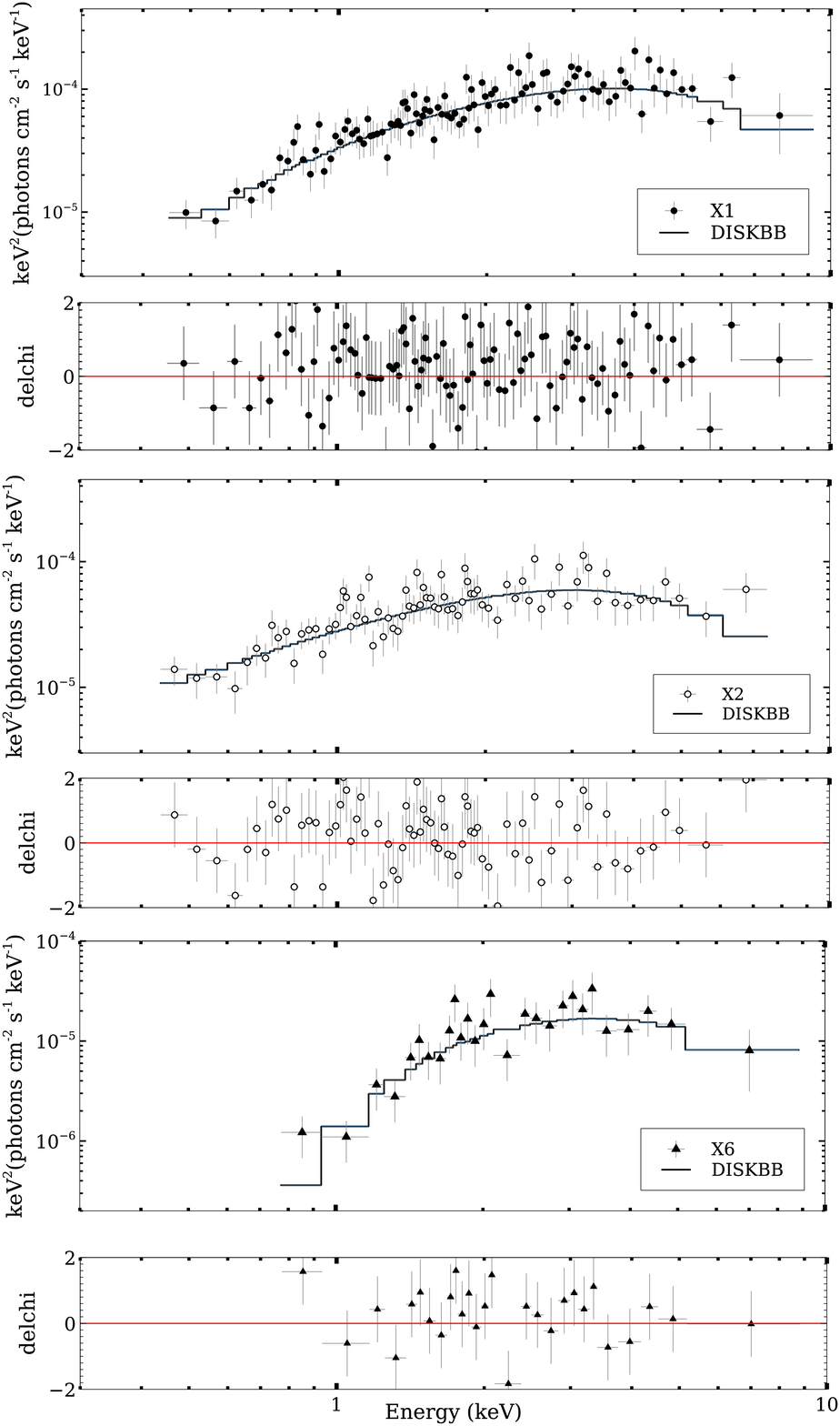}
\caption{Unfolded energy spectra of X1 (upper), X2 (center) and X6 (lower) by using {\it Chandra} data in the 0.3-10 keV energy range. The units of delchi is (data-model)/error.}
\label{F:xspec}
\end{center}
\end{figure}

In Table \ref{T:xmodel}, the well$-$fitted spectral model parameters for X1, X2 and X6 (deploying the {\it Chandra} data only) are given. 
 The acceptable model-fits were selected according to null hypothesis probability. We consider all fits with P $>$ 0.05 corresponding to a confidence level >95\%.
 We found the {\it diskbb} model fits to be statistically better than the other single component models at the 3$\sigma$ confidence level. The unfolded energy spectra of the best model-fits, for X1, X2 and X6, are shown in Figure \ref{F:xspec}. The ULX candidates with low data statistics (i.e X5, X7, X8 and X9) were also fitted with the same single component models using C-statistics. However, none of the models gave reasonable results.
 
 \begin{table*}
\centering
\begin{minipage}[b]{0.9\linewidth}
\caption{The {\it diskbb} model parameters of three ULXs from the {\it Chandra} observation}
\begin{tabular}{c c c c c c c r r r l }
\hline
Source & N$_{\mathrm{H}}$ & N$_{\mathrm{{\it diskbb}}}$ & T$_{\mathrm{in}}$ & F$_{\mathrm{X_{unabs}}}$ & L$_{\mathrm{X}}$ & {\it P}\\
(1) & (2) & (3) & (4) & (5) & (6) & (7)\\
\hline
X1 & $0.08_{-0.02}^{+0.02}$ & $2.94_{-1.56}^{+1.57}$ & $1.50_{-0.02}^{+0.02}$ & $3.10_{-0.14}^{+0.14}$ & $9.84_{-0.46}^{+0.46}$  & 0.73 \\
X2 & $0.01_{-0.01}^{+0.02}$ & $3.30_{-1.78}^{+1.78}$ & $1.28_{-0.02}^{+0.02}$ & $1.81_{-0.10}^{+0.10}$ & $5.76_{-0.31}^{+0.31}$  & 0.63\\
X6 & $0.86_{-0.16}^{+0.22}$ & $1.07_{-2.82}^{+2.82}$ & $1.25_{-0.05}^{+0.05}$ & $0.57_{-0.08}^{+0.08}$ & $1.82_{-0.27}^{+0.27}$  & 0.76\\
\hline
\end{tabular}
\\ Note. — Col. (1): Source label. Col. (2): X-ray absorption value, in units of 10$^{22}cm^{-2}$. Col. (3): Normalization parameter of {\it diskbb} model; 10$^{-3}$$\times$([(r$_{in}$ km$^{-1}$)/(D/10 kpc)]$^{2}$ $\times cosi$). Col. (4): Temperature at inner disk radius (keV). Col. (5): Unabsorbed flux in units of 10$^{-13}$ ergs $cm^{-2}$ $s^{-1}$.
Col. (6): Unabsorbed luminosity in units of 10$^{39}$ ergs $s^{-1}$. Col. (7): The null hypothesis probability, which is the probability of the observed data being drawn from the model given the value of $\chi^2$ and the dof. The reduced $\chi^{2}$ is given in parentheses. All errors are at 90\% confidence level. Unabsorbed flux and luminosity values are calculated in the 0.3–10 keV energy band. Adopted distance of 16.3 Mpc \citep{2000AJ....119..612D} is used for luminosity. \\
\label{T:xmodel}
\end{minipage}
\end{table*}

\begin{figure}
\begin{center}
\includegraphics[width=\columnwidth]{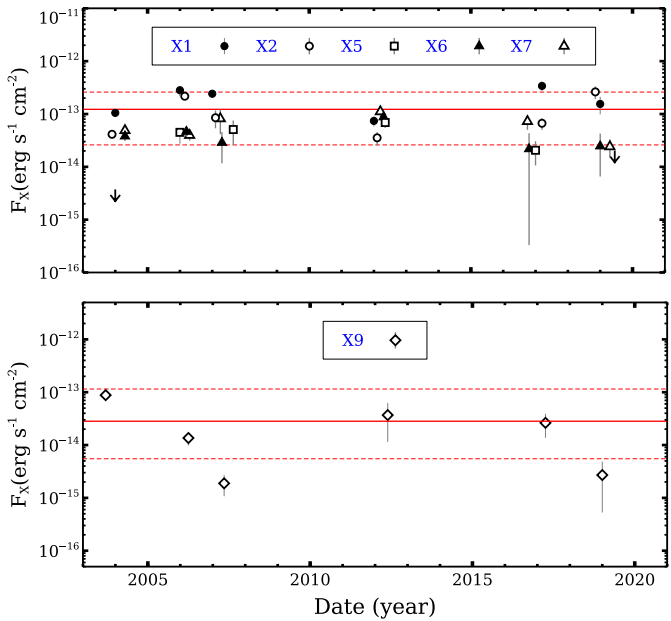}
\caption{Upper: The long-term X-ray light curve for the ULX candidates which have optical counterparts. The vertical arrows represent 3$\sigma$ upper limits when the source is not detected. Lower: The long-term X-ray light curve of X9. In both panels the solid line indicates the mean value of the X-ray flux; a measure of the variability is provided by the $\pm$ 3$\sigma$ levels shown as dotted lines. Observations are from {\it XMM-Newton} (2004), {\it Chandra} (2006) and {\it Swift-XRT} (2007; 2012;2017 and 2019).}
\label{F:lcxrays}
\end{center}
\end{figure}

We used all the available X-ray data to examine the long-term and sort-term variability of the candidate ULXs. The light curves of the sources which have optical counterparts are shown plotted in Figure \ref{F:lcxrays}. The source X9 seems to exhibit more variability than the others. We also performed a periodicity search for all of the candidate ULXs except X4. Our results showed no evidence for any significant periodicity $\geq$ 1.5$\sigma$ in any of the sources examined.

\begin{figure}
\begin{center}
\includegraphics[width=\columnwidth]{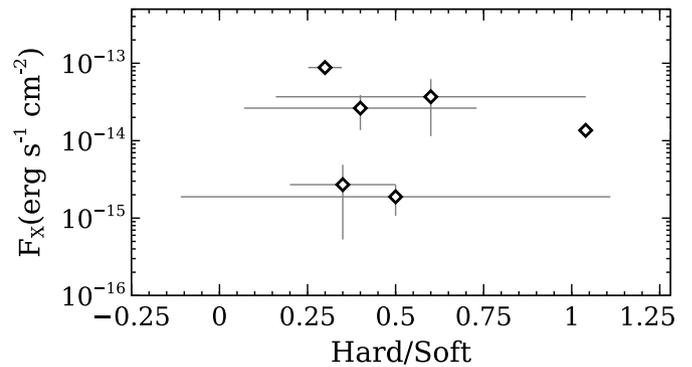}
\caption{The HID for X9 suggestive of a partial q-curve pattern.}
\label{F:HIDX9}
\end{center}
\end{figure}

In order to probe possible state transitions in the ULXs, we constructed hardness-intensity diagrams (HIDs) and hardness ratio diagrams (HRDs) for the sources that have optical counterparts. The HRs are calculated as the ratio of counts in the hard-to-soft energy bands where the respective bands are as follows: 0.3 - 2.0 keV (Soft) and 2.0 - 10.0 keV (Hard). The HID of X9 are displayed in Figure \ref{F:HIDX9}. No significant results were obtained for remaining sources.

\section{Discussion and CONCLUSIONS} \label{sec:3}

A recent multiwavelength study of NGC 1672 identified multiple bright X-ray sources including 9 ULX candidates. In this study, we performed an optical, spectral, and a temporal analysis of these ULX candidates. We deployed archival optical data from {\it HST}, and X-ray data from {\it Chandra}, {\it XMM-Newton}, and {\it Swift-XRT}. The precise source positions were obtained through improved astrometry based on the {\it Chandra} and {\it HST} observations to search for and identify potential optical counterparts for the ULX candidates. We summarize our main findings as follows:

\begin{enumerate}

\item We identified unique optical counterparts for X2 and X6.

\item In the case of X1, X5 and X7, we were able to isolate two optical counterparts for each source.

\item No optical counterparts were found for X3, X8 and X4 within the respective error circles. The source X9, unfortunately did not have any {\it HST} data, so it could not be investigated further.

\item The absolute magnitudes ($-$5 <M$_{V}$< $-$7.5) and spectral types (B-A) of the identified optical counterparts in NGC 1672 are compatible with optical companions of ULXs in other galaxies; see \cite{2013ApJS..206...14G} and \citealp{2011ApJ...737...81T}. Moreover, the optical counterparts appear to be faint in the V$-$band and relatively brighter in the UV band.

\item The SED for the counterpart {\it X1$\_1$} is well-fitted with {\it power-law}, F$\propto$ $\lambda^{\alpha}$, with a photon index, $\alpha$$=$$-$ 2.13 and also shows 0.6 mag. variability in UV. These findings are consistent with optical emission arising primarily from an accretion disk. Furthermore, the SED of {\it X2$\_1$} is adequately fitted with a {\it blackbody} model and a temperature of 10$^{4}$ K. In this case, the emission could come from the donor star and/or result from irradiation of the disk.

\item The sources X5 and X9 exhibit high X-ray variability, with the flux varying by factor of $\sim$ 30 and 50 times, respectively, over observations spanning a time period of 2004 to 2019. Due to the sparse coverage of the data, it is difficult to interpret that the high variability exhibits bi-modal flux distribution that could indicate the propeller effect. This level of variability together with the fact that X9 displays a partial q-curve track in the hardness-intensity diagram (possibly indicating spectral transitions) makes it an interesting source for further investigation.
\end{enumerate}

\section*{Acknowledgments}

This research was supported by the Scientific and Technological Research Council of Turkey (TÜBİTAK) through project number 119F315. ES and KD acknowledge support provided by the TÜBİTAK through project number 119F334

\bibliography{Wiley-ASNA}

\end{document}